\documentclass[letterpaper]{article} 
\usepackage[]{aaai25}  
\usepackage{times}  
\usepackage{helvet}  
\usepackage{courier}  
\usepackage[hyphens]{url}  
\usepackage{graphicx} 
\usepackage{natbib}  
\usepackage{caption} 
\usepackage{enumitem}
\usepackage{algorithm}
\usepackage{algorithmic}
\usepackage{xcolor}
\usepackage{amsmath}
\newcommand{\answerYes}[1]{\textcolor{blue}{#1}} 
 
\newcommand{\answerNA}[1]{\textcolor{gray}{#1}}

\newcommand{\xhdr}[1]{\vspace{1.7mm}\noindent{{\bf #1.}}}
\newcommand{\xhdrNoPeriod}[1]{\vspace{1.7mm}\noindent{{\bf #1}}}

\usepackage{soul}
\setuldepth{strikethrough} 

\usepackage{circledsteps}
\newcommand\myCircled[2][]{\ifmmode
\Circled[fill color=black,inner color=white,#1]{\mathsf{#2}}
\else
\Circled[fill color=black,inner color=white,#1]{\sffamily#2}
\fi
}

\usepackage{newfloat}
\usepackage{listings}
\DeclareCaptionStyle{ruled}{labelfont=normalfont,labelsep=colon,strut=off} 
\floatstyle{ruled}
\newfloat{listing}{tb}{lst}{}
\floatname{listing}{Listing}

\usepackage{listings}
\usepackage{xcolor}
\usepackage{xtab}
\usepackage{booktabs} 

\lstdefinestyle{prompt}{
  basicstyle=\ttfamily\tiny,
  frame=single,
  breaklines=true,
  showstringspaces=false,
  captionpos=b,
  numbers=none,
  aboveskip=10pt, 
  belowskip=10pt,
}

\title{Alexandria: A Library of Pluralistic Values for \\Realtime Re-Ranking of Social Media Feeds}
\author{
    Akaash Kolluri\equalcontrib\textsuperscript{\rm 1},
    Renn Su\equalcontrib\textsuperscript{\rm 1}, 
    Farnaz Jahanbakhsh\textsuperscript{\rm 2}, 
    Dora Zhao\textsuperscript{\rm 1}, \\
    Tiziano Piccardi\textsuperscript{\rm 1}, 
    Michael S. Bernstein\textsuperscript{\rm 1}
}
\affiliations{

    \textsuperscript{\rm 1} Stanford University
     \textsuperscript{\rm 2} University of Michigan \\
     farnaz@umich.edu, \{dorothyz, mbernst\}@stanford.edu
}

\usepackage{bibentry}

\begin{document}

\maketitle

\begin{abstract}
Social media feed ranking algorithms fail when they too narrowly focus on engagement as their objective. The literature has asserted a wide variety of values that these algorithms should account for as well---ranging from well-being to productive discourse---far more than can be encapsulated by a single topic or theory. In response, we present a \textit{library of values} for social media algorithms: a pluralistic set of 78 values as articulated across the literature, implemented into LLM-powered content classifiers that can be installed individually or in combination for real-time re-ranking of social media feeds. We investigate this approach by developing a browser extension, \textit{Alexandria}, that re-ranks the X/Twitter feed in real time based on the user's desired values. Through two user studies, both qualitative ($N$=12) and quantitative ($N$=257), we found that diverse user needs require a large library of values, enabling more nuanced preferences and greater user control.
With this work, we argue that the values criticized as missing from social media ranking algorithms can be operationalized and deployed today through end-user tools.
\end{abstract}

\section{Introduction}

Which values should shape the behavior of our social media feed ranking algorithms? Today, most algorithms optimize for engagement signals such as likes and reshares~\cite{backstrom2016serving,narayanan2023}.
While effective at maximizing user interactions, engagement-based ranking prioritizes a narrow set of platform-centric objectives and specific individualistic values~\cite{bernstein2023embedding}, often neglecting broader societal values. Reward hacking on these values can create problematic trade-offs~\cite{fourcade2024ordinal}, for example, harming well-being~\cite{orben2022windows,burke2020social}, reinforcing marginalization~\cite{marshall2021algorithmic}, and threatening healthy democratic discourse~\cite{jia2023embedding,lorenz2023systematic}. Platforms have implemented mitigations that integrate non-engagement outcomes~\cite{eckles2022algorithmic,eckles2016estimating} or surveys to obtain more reflective feedback from users~\cite{backstrom2015}, but so far, these mitigations have not fully addressed the issues at stake.

In response to this challenge, researchers and practitioners have proposed a series of alternative feed algorithms that address these shortcomings. Proposed interventions, for example, might promote posts that are more supportive of a healthy democracy~\cite{bail2022breaking,jia2023embedding,piccardi2024social}, they might promote content that is less likely to harm well-being~\cite{stray2020aligning}, or they might downrank suspected misinformation~\cite{fb_enforcement, twitter_enforcement, youtube_enforcement,bhargava2019gobo}.
Each of these interventions is typically positioned as an intervention to align the feed with a single societal value---an intervention to address a specific gap in feed behavior. However, the social sciences assert the importance of not just single values, but of representing an incredibly wide variety of values~\cite{tetlock1986value,rokeach1973nature}. 
To succeed, feed algorithms need to balance across many competing values, not optimizing one at a time. Nor does any single set of universal values apply to an entire community or platform~\cite{zhu2018value,weld2022makes}.

We present Alexandria, an extensible library of 78 individual and community values that lets social media users customize their own feed ranking by prioritizing the values they want to see more in their feed. Named after the historical Great Library of Alexandria, a library renowned for its wide-ranging curation of human knowledge across the ancient world, our system similarly aims to enable a pluralistic approach to value-embedded feed curation.
We bootstrap this library with a set of over 100 values\footnote{In this context, the term \textit{values} encompasses any human, community, or societal value that can be expressed in social media posts and that users may want to customize.} generated by combining the value systems proposed by six different review papers~\cite{rokeach1973nature,hofstede1984culture,weld2022makes,stray2022building,maslow1987maslow,ge2018taxonomy}, each nominating values that social media algorithms ought to optimize. We translate each of these source values into a feed ranking objective using a large language model (LLM), adapting a method that has resulted in high agreement with expert annotation~\cite{jia2023embedding}. We then remove correlated, overlapping values, obtaining our library of 78 values. 
We instantiated this library of values into a browser extension compatible with X/Twitter that allows end users to select values to up- or down-rank in their own feeds. Alexandria weights the user-selected values and re-ranks the feed in real time according to the selected values for the end user.

We conduct two studies to investigate the potentials of the library and how users interact with it. In the first study, we interview participants (N=12) who use the Alexandria browser extension on their X/Twitter feeds. Participants appreciate the control that the tool provides and report an increased awareness of the values underlying their feeds. Participants find that the full library, compared to a single value system, makes it easier to identify values that match their needs. To examine patterns of library use at scale, we then run a controlled experiment where participants (N=257) have access to either the full library of values or a set of values originating from a single value system. Participants are asked to re-rank their X/Twitter feed using the provided values settings in Alexandria. We find that users can articulate their desired values more precisely when given access to the full library. While there exist values that are commonly used across conditions, the diversity of values selected by our participants reinforces the benefit of a library containing specific and more nuanced values.

This work can inform new platform designs---supporting the trend of custom feeds design and algorithmic choice in decentralized platforms~\cite{kleppmann2024bluesky}. While prior work has argued for and operationalized re-ranking feeds based on values, the modeling thus far has focused on single values selected at the discretion of researchers or platforms, with no input from end-users~\cite{jia2023embedding, bernstein2023embedding}. What remains underexplored is: (1) how to embed a plurality of values into feeds, (2) how to incorporate user input into the process, and (3) how to implement the approach in a way that supports scalable, real-time re-ranking of feeds. Specifically, we contribute:

\begin{itemize}[topsep=0pt,itemsep=-1ex,partopsep=1ex,parsep=1ex]
    \item A sample library of values for social media feeds, drawn by combining multiple previous frameworks that emphasize different subsets of values
    \item A generalizable operationalization pipeline for translating any set of value systems into feed ranking objectives
    \item Our method for enabling value pluralism in ranking social media feeds is instantiated in Alexandria---a browser extension that allows users to explore, select, and apply values to their X/Twitter feeds in real-time
    \item An empirical understanding of how users perceive and use the system on their own feed, and an evaluation demonstrating that our library provides greater coverage of the values that participants find useful for ranking feeds compared to single value system alternatives
\end{itemize}

\section{Related Work}
\xhdr{Values in Social Media Feeds}
Today, feed ranking systems focus on metrics of user engagement, which serve as proxies for user satisfaction and, ultimately, platform revenue~\cite{eckles_2022,milli2021optimizing,ciampaglia2018algorithmic, twitteralgorithm}. These algorithms can end up maximizing individual experience at the cost of societal values such as democracy or political tolerance. For instance, maximizing engagement can amplify anti-social behaviors such as online harassment ~\cite{are2020instagram, munn2020angry} or focus our attention on pro-attitudinal political content ~\cite{sunstein2001http,rowland2011filter}. However, in contrast to the status quo, a growing body of literature argues for prioritizing human-centered values in recommender systems~\cite{bernstein2023embedding,stray2020aligning,stray2022building}.

Social media platforms themselves have offered methods for incorporating some notions of societal values into ranking models. Content moderation models are a common strategy to ensure that content does not violate platform policies or community guidelines~\cite{gillespie2018custodians}. Beyond content moderation, platforms have also taken measures to combat the societal harms posed by such content as extremism, misinformation, and violence with automated detection and removal~\cite{fb_enforcement,twitter_enforcement,youtube_enforcement}. 

Researchers have also investigated interventions on social media feeds to counter user harms relating to societal values. For example, allowing users to aggregate and filter content across platforms~\cite{bhargava2019gobo} and enabling users to filter out inaccurate content~\cite{jahanbakhsh2022leveraging, jahanbakhsh2024browser}. Other work has explored interventions that expose users to alternative feeds, such as contrasting liberal and conservative on Facebook~\cite{bluefeed_redfeed} or Twitter~\cite{saveski2021perspective}, re-ranking feeds to achieve ideological balance~\cite{celis2019controlling, babaei2018purple} or build trust across divides~\cite{ovadya2023bridging}. 
Studies on Facebook have also examined the effects of reducing exposure to content from like-minded sources~\cite{nyhan2023like, guess2023social}, reshared content~\cite{guess2023reshares} or substituting the algorithmic feed with a reverse-chronologically-sorted feed~\cite{guess2023social} during the 2020 US presidential election.

Frameworks from the social sciences are useful for identifying values that might be integrated productively into these algorithms. Using large language models (LLMs), researchers can efficiently label constructs from the social sciences in social media content---such as support for democratic principles---thereby allowing them to directly encode these concepts into the ranking algorithm~\cite{jia2023embedding,bernstein2023embedding,piccardi2024social}.

\xhdr{User Control on Social Media}
On social media platforms, users often lack control over their feeds. In the absence of explicit controls, users attempt to make sense of the complex and opaque algorithms, developing informal explanations or ``folk theories'' of how these algorithms function~\cite{eslami2015always}. 

Users leverage these folk theories to guide how they interact with content on the platform. For example, to communicate with the algorithm not to show them certain content, users may like other content that they, in fact, have no strong preference for~\cite{eslami2016first,jahanbakhsh2022leveraging}. In addition to these informal techniques that provide uncertain efficacy, platforms themselves can introduce better feed controls for users~\cite{feng2024mapping}. For instance, on X/Twitter, users are able to specify ``muted words'' that they do not want to see on their timeline~\cite{twittermuted}. Platforms, such as X/Twitter and Instagram, also give users the option to opt out of the algorithmic ranking and instead see content in chronological order. On the other hand, TikTok attempts to enhance transparency by providing explanations for its recommendations, though these explanations do not always align with the user’s actual behavior~\cite{mousavi2024auditing}. Recent attention to users' desire for greater control over their data and feed content has contributed to the development of alternative social media platforms with dedicated decentralized protocols, such as Bluesky and Mastodon~\cite{kleppmann2024bluesky}, that support personalization and custom feeds.
However, in most centralized platforms, user control and transparency are still limited to constrained platform settings or must be navigated through informal means.

To address this issue, a body of work has sought to design explicit end-user controls for social media feeds. One way to instantiate these controls is through new user interface designs~\cite{dooms2014improving,wang2024end,ekstrand2015letting, knijnenburg2011each, jannach2017user,friedman2023leveraging, jahanbakhsh2023exploring}. Proposed solutions include presenting different options, such as word-level filters, sliders, and toggles, that can be used to configure personalized content moderation settings~\cite{jhaver2023personalizing}, introducing explicit ways to teach the ranking algorithm the desired behavior~\cite{feng2024mapping}, and offering more control to the user to break the filter bubbles~\cite{liu2024does}.

\section{Creating a Library of Values}

While prior work has argued for ranking social media feeds using societal values, the question remains: what values do we choose? Rather than selecting one value system, our goal is to design a flexible method that allows us to draw from a plurality of value systems for ranking feeds. 
Our method is robust enough to operationalize \emph{any} value system.

\subsection{Defining Value Systems}
Our work adapts existing value systems for social media feed ranking. We define \textit{value systems} as established methods of value categorization drawn from academic literature. These systems provide a method for articulating abstract beliefs, morals, and ideals of both individuals and collectives~\cite{rokeach1973nature,schwartz1987toward,graham2013moral,kluckhohn1951values}. In this paper, we make no claim that the value systems we select are the ``best'' in some absolute sense---instead, we select well-known value systems as proofs of the generalizability of our concept. We demonstrate a procedure that can integrate other values or value systems as they are proposed. To this end, we populate our library with both general-purpose value systems from the social sciences as well as domain-specific value systems for social media. In the prior category, we use three classical value systems from psychology: Rokeach's Value Survey~\cite{rokeach1973nature}, Maslow's hierarchy of needs~\cite{maslow1987maslow}, and Hofstede's cultural dimensions~\cite{hofstede1984culture}. These value systems have been extensively validated and shown to be comprehensive across cultural and temporal contexts~\cite{schwartz2012overview}. In addition, we also consider more bespoke taxonomies that focus on domains similar to this work: social media and recommender systems. We choose the following three taxonomies: \citet{stray2022building}'s set of 30 values for recommender systems; \citet{weld2022makes}'s taxonomy of community values on Reddit; and \citet{ge2018taxonomy}'s co-creation values on Weibo. These systems introduce values that capture unique features of social media feeds unrelated to basic human values, such as humor or prioritizing content quality.

\subsection{Operationalization framework}
Here, we describe our method for constructing a library of values given a set of existing value systems. This process (Fig.~\ref{fig:library}) consists of three steps: identifying relevant values for re-ranking, adapting the definitions to be used by an LLM, and then merging overlapping values. 

\xhdr{Identifying Usable Post-Level Values}
The first step is to determine the relevant values that are germane for ranking feeds at the \textit{post} level. To identify the set of values, we manually reviewed all 145 values and their associated definition provided by the value systems' authors. From this initial pool, we filtered out values that did not meet the following criteria: (1)~Most importantly, the value must refer to the content of a post. For example, a value like ``Compassion'' can be expressed in the content of a post, but the value of ``Control over your recommender'' is a platform-level statement that does not apply to a single post. (2)~Second, the value must be identifiable from individual post content. Values such as ``Diversity of Content'' are excluded in this step. (3)~Finally, we exclude values whose content classifications highly depend on individual subjectivity (e.g., ``Personal Preferences''), as these objectives can be incorporated by classic personalization or engagement-based models. Two different research team members applied these three criteria individually and then met to resolve any differences. This process left 111 values.

\xhdr{Translating Value Definitions}
After identifying our set of values, we next operationalized each construct so that it could be used by an LLM for labeling social media content. We aim to create classifiers that determine the magnitude of each value expressed in a social media post, categorizing it as \textit{not present}, \textit{weakly present}, or \textit{strongly present}.

First, we convert the value textual definitions into a standardized format by removing references to a specific platform and irrelevant details (i.e., citations or platform-specific terminology). Already usable definitions are not modified. Furthermore, we do not change the specific phrasing and level of abstraction of provided definitions. Our goal is to preserve authorial intent while keeping the definition platform agnostic and self-contained to a post.

As an example, \citet{ge2018taxonomy} identify ``Caring for Others'' as one of the core values within Weibo communities. In their paper, this value is defined as follows: ``Relating to destination marketing organizations (DMO)-published non-product-related posts, users express their concerns for others such as their health and happiness.'' When translating for our use case, we remove language surrounding ``DMO-published'' and ``non-product-related,'' distilling the definition to ``users express their concerns for others such as their health and happiness.''

Once we have our standardized definitions, we then import each one into a labeling prompt for LLM classification (Listing~\ref{lst:prompt}). Prior work has established that LLMs accurately code the expression of social science constructs, achieving as similar inter-rater agreement rates with experts as the experts achieve with each other~\cite{ziems2023large,jia2023embedding}.

\xhdr{Evaluating Model Performance} We validated the LLM's performance on our task by comparing the generated labels against human annotations. From each value system, we randomly selected two values (i.e., 12 values in total) and then sampled 30 posts per value from the datasets generated in our user studies reported later. For each value, the posts were labeled by 5 annotators; we use the mean over the ratings rounded to the nearest integer as our ground-truth label. When comparing the human and generated labels, we use two metrics: binary accuracy when classifying whether the value is present (i.e., label = 0) or absent (i.e., label $>$ 0) in the post and mean absolute error (MAE).

Binary accuracy is 81.2\%, indicating that the classifier can identify whether a value is present or not in posts. This result is robust across values, with accuracy ranging from $68.8\%$ to $100.0\%$. MAE, which can range from 0 to 2, is $0.45$, indicating close alignment with human annotations. Across values, MAE ranges between $0.28$ (for ``Biological and Physiological Needs'') and $0.75$ (for ``Restraint''). Even the highest MAE of $0.75$ is less than $1$, suggesting that, on average, the LLM's predictions are reasonably close to the ground truth, only occasionally deviating by an entire category (e.g., predicting \emph{weakly present} instead of \emph{not present}). We report these metrics, as well as recall and F1, disaggregated by individual values in the Appendix. 

Furthermore, we compared our classifier's performance against the disagreement between human raters. For each value, we calculated the mean absolute error between each human vote and the average of the other four votes. The LLM's MAE ($0.45$) outperforms human votes ($0.61$), or in other words, is closer to the average human label; this pattern is consistent across 10 of the 12 values. These results suggest that our classifier performs as well, if not better, than a single human annotator at estimating the strength of a value in a post as perceived by average humans. We perform additional analyses to validate our value-ranking pipeline. See Appendix for more details on the evaluation of the classifier as well as the entire pipeline.

\xhdr{Merging Similar Values}
Finally, since different value systems may contain overlapping constructs, we merge similar values together to reduce redundancy. To identify which values tend to co-occur frequently, which can indicate overlapping constructs, we analyzed a sample of tweets sourced from users' real feeds of $100$ unique X/Twitter users. For each of them, we collected approximately 480 posts from their ``For You'' page, labeled them with the LLM classifiers for the presence of the $111$ values, and sampled 100 posts from each value.
Then, we calculated Pearson's correlation for value occurrence across tweets for every combination of constructs. We merge pairs of constructs with coefficients $r\geq0.6$. 

We selected $0.6$ as our threshold for merging constructs, as this is typically considered the upper bound of a moderate correlation~\cite{schober2018correlation}, but the merging threshold may be changed based on the desired application or might allow for human oversight and manual merging. While our procedure may end up merging values that one might argue are distinct, there is a direct trade-off between the manageability of the library size for users and the granularity of values. In this work, we use a greedy algorithm, iterating through all pairs of constructs. If the correlation exceeds our threshold, we merge the pair, retaining the first value. After this process, $78$ values remained, forming our library of values. Overall, we retain over $50\%$ of all original values and all removed were either not operationalizable through our means or strongly correlated with another value from our system. See Table~\ref{tab:value-systems} for details on the number of values retained after filtering and merging. Further, we note that our library maintains extensive coverage of tweets on Twitter—fewer than 3\% of the tweets we labeled during merging contain zero values in our final library, indicating that our values are broad enough to rank nearly all content on Twitter (a graph breaking down the distribution of values per Tweet is available in the Appendix \ref{appendix:coverage}).

\section{Alexandria: Reranking Social Media Feeds Using a Library of Values}

To apply our library of values for re-ranking actual feeds, we introduce Alexandria---a browser extension for X/Twitter that allows users to re-rank their feeds in real-time based on the values they want to personalize. In this section, we describe Alexandria's design, walking through how users can efficiently search for values and apply them to re-order their own social media content.

\subsection{System Design}

\begin{figure}
    \centering
    \includegraphics[width=1.0\linewidth]{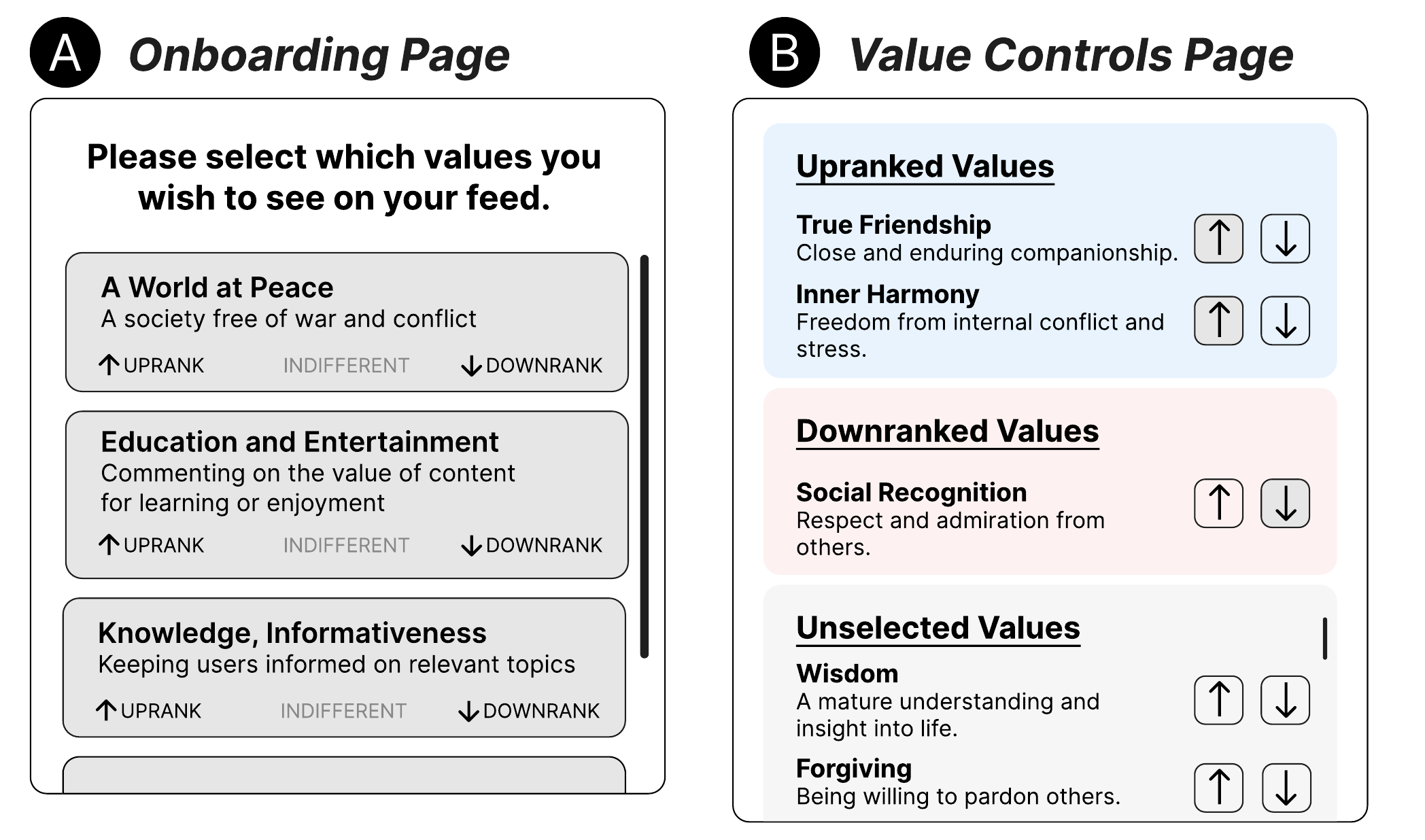}
    \caption{After selecting the values they wish to see from an initial set of five, users can then iteratively refine their feed by upranking or downranking additional values. The Onboarding Page \myCircled{A} is shown to users when they first install the extension. At any time, users can refine the values they selected using the Value Controls Page \myCircled{B}.}
    \label{fig:extension}
\end{figure}

Alexandria consists of two main components: (1)~\emph{value selection} (e.g., the user must select the values from the library that they want to prioritize or downrank in their feed), and (2)~\emph{realtime feed re-ranking} (e.g., the feeds must be re-ordered in real-time in the browser).

\xhdr{Value Discovery and Selection}
\label{sec:onboarding}
Given the large set of values, we build a search and recommendation system to make exploring the value library more manageable for users. While users can fine-tune and revise their choices at any point, our first step is to propose a diverse set of values that users can use to initialize the web extension. After the installation the extension opens the \myCircled{A}~\emph{Onboarding Page} in the popup pane and shows the user a representative set of five values---``A World at Peace'', ``Education and Entertainment'', ``Knowledge, Informativeness'', ``Appreciation'', and ``Collectivism''. 
We curated this set of five by first embedding the name and definition using OpenAI and then clustering the respective vectors with K-Means. For each cluster, we identified the value most similar to its centroid (using cosine similarity) and used these values to seed our recommender system. When presented with this set, users have the option to uprank, downrank, or not rank each of the values. This step is to ensure that the initial set of values encode diverse concepts and are representative of the available options. See Fig.~\ref{fig:extension} for a visualization of the interface.

After the initial selection, we dynamically recommend additional values for the user. We recommend modules that are similar to those that the user configured in the onboarding step. To find similar values, we compute the vector embedding representation of the set of values the user has upranked or downranked. We represent the preference as an average of the embeddings for the selected values multiplied by the weight a user places on the value (1 if the value is upranked and -1 if the value is downranked). Then, we sort the remaining values in our library by the cosine similarity between the preference vector and the embeddings of each of the other values in the library. We present the top ten of the sorted values to our users. Again, they have the option to uprank, downrank, or ignore each value. 

Finally, users can refine their choices by either changing the values selected or adjusting the weights of the values. To explore all values, users can navigate to the \myCircled{B}~\emph{Value Controls Page} in the extension's popup. This page consists of a list of values and their corresponding definitions, presented in three panels (Fig.~\ref{fig:extension}): the already upranked values, the already downranked values, and an expandable list of values that have not yet been ranked. To change the strength of each rating, users can adjust sliders corresponding to the weights placed on each value. The strength applied to upranking and downranking can vary from $0.1$ to $1$.

\xhdr{Real time Feed Reranking}
The next component of Alexandria is the feed re-ranking. We integrate our browser extension directly into the X/Twitter platform. To keep the platform experience consistent for our users, the re-ranking needs to be performed in as close to real-time as possible and working with content from the user's actual feed without disrupting the user experience~\cite{piccardi2024reranking}. We detail the process for obtaining the user's posts and then re-ranking them based on values as follows. 

First, to gather the posts for re-ranking, the extension silently simulates scrolling through the ``For You'' timeline before displaying the feed, collecting approximately $70$ posts. These posts represent the initial inventory. 
Note that the extension works on the user's algorithmically curated feed, meaning the inventory contains tweets that have already been filtered for engagement. 

Next, to rerank the feed by user-configured values, we send each tweet to our own server, where we use \texttt{GPT-4o-mini} to label the presence and strength of each value that the user has activated (Listing~\ref{lst:prompt}). The rating is on a three-point scale: 0 signifies the value is not reflected; 1 signifies the value is weakly reflected; and 2 signifies the value is strongly reflected. The labeling is performed on the text of the tweet and any associated multimedia content (e.g., embedded image, link metadata). 

Finally, once we obtain the value labels, we determine the ranking order by computing, for each post, the dot product between the rating vector estimated by the LLM and the user's chosen weight for each value:
\[
s_i = \mathbf{r}_i \cdot \mathbf{w} = \sum_{v=1}^{V} r_{i,v} w_v
\]
where $r_{i,v}$ is the LLM estimation for the strength of value $v$ in post $i$ (with values in $\{0,1,2\}$), $w_v$ is the user-configured weight for value $v$, and $s_i$ is the final score of the post $i$.
Intuitively, the dot product is a measure of the alignment of user-tweet values. The tweets are then sorted in descending order based on the score from this dot product. The entire process takes, on average, from five to ten seconds---adding a small latency to the load experience, much of which is waiting for X/Twitter to return enough content to rerank.

This process repeats each time the user triggers the infinite scroll to see more tweets---whether by scrolling down or clicking on the floating icon at the top of their screen---or when they update their value selections in Alexandria.

\section{Qualitative User Study}
We conducted a qualitative user study of our tool to evaluate whether i) our real-time re-ranking system provides value to users and ii) the proposed library of values provides a benefit over a single value system. 

\subsection{Method}

We recruited 12 participants via Prolific and posting on X/Twitter for a 30-45 minute interview. To qualify, participants must be 18 years or older, US citizens, have an X/Twitter account, and use Google Chrome. We compensated participants with \$20. See Table~\ref{tab:demographics} for participant demographics.

\xhdr{Study Procedure} First, participants described their current social media use, including what they currently like and dislike about content recommended to them and how they currently try to assert control over the content they see. Next, participants installed the Alexandria extension where they were initially randomly assigned a single value system, without knowledge that a larger library existed. Participants were asked to select values and iteratively rerank their feed. We asked about how satisfied they were with their crafted feed, what values they liked, and what values they were missing. Then, participants were given access to the full system of values, and repeated the same process for the single system. Finally, they were asked to compare their experiences using the two systems. 

\xhdr{Analysis} To categorize the themes that surfaced in users' qualitative data related to our questions, a member of the research team conducted initial coding by making multiple passes over the interview scripts, using an approach inspired by grounded theory to assign codes to the idea units. The codes and themes were subsequently refined through discussions within the research team~\cite{muller2012grounded}.

\subsection{Benefits of Alexandria for Feed Re-ranking}
\xhdr{Alexandria addresses user frustrations around lack of control}
Many participants reported that Alexandria helped them overcome their frustrations with the lack of control over their social media feeds. These frustrations included feeling that their feeds displayed content that seemed arbitrary, deviated from their preferences, or that was forced on them despite their explicit aversion. Participants also mentioned that current recommendations were over-optimized for their engagement, which left them unhappy with how they spent their time: \textit{``I try to just stay within the people I'm following, but I do often find myself navigating over to like a For You page or like the Explore page, which can often be a time drain. (P11)''} This observation suggests that the myopic, engagement-driven design of current feeds is at odds with users’ aspirational goals and sense of intentionality---tensions that Alexandria helps address by making feed values visible and adjustable.
Even when features were available for feed curation on a platform, such as the option to mark ``Not Interested'' on content, participants who used them noted limited efficacy as changes from these features were not durable. As P11 explains, \textit{``if you make the wrong turn... [you] wind back up in that same topic space.''}

Our reranking system weaves users' value-driven priorities into engagement-based rankings, giving users greater control over the kinds of content they wish to see while maintaining the relevance of engagement-focused feeds. Participants expressed satisfaction with the control they found through the iterative process of reranking their feed and observing the immediate outcome: 
\textit{``I felt like I definitely did have control. [I] definitely saw results from updating and re-ranking my feed based on the changing the values ... the first time I ranked my values using the full version, I did see a lot of like political [content]. After re-ranking the values ... [I thought] 'What values did I select that might be causing that? .. What values should I like explicitly downvote to avoid seeing that content?' ... After going through that process ... I did notice a big difference.
(P12)''}

\xhdr{Alexandria's value controls help align feeds to users' content preferences} Participants expressed clear preferences for specific types of content on their feeds, articulated in terms of values, topics, or stylistic features. Even when not explicitly framed as values, many of these preferences reflected underlying value preferences. Participants found that selecting values in Alexandria helped bring their feeds into closer alignment with these preferences. For example, P7 was \textit{``interested in national security, family security, and I like people that have wise things to say.''} Yet, even when specific values were not mentioned, participants' preferences often aligned with certain underlying values. For example, P4 enjoyed\textit{``seeing stand-up bits,''} mapping to the value of ``Adding humor''. P11 shared that they ``\emph{follow a lot of professors and people in my field on X}'', which reflects values such as ``Knowledgeable people'' or ``Wisdom''. By giving users control over values, in contrast to other mappings, such as topic or sentiment, Alexandria provides a way to shape their feed that is broad enough to capture diverse preferences and focused enough to remain practical and intuitive.

\xhdr{Alexandria promotes value awareness} Before using the tool, many participants did not have specific values in mind that they wanted on their feed. However, after exploring the available values and observing the outcome of the feed reranking, some participants reflected on how the types of content they saw related to values. This led to participants to become more aware of what values they wanted to see or avoid on their feed: \textit{``There's one [value] here. It says obedient, and I'm thinking of those trad wife videos, and I'm like, that's not for me. (P8)''}

\xhdr{The full library offers more flexibility in value selection}
After using both systems, most participants (10 of 12) said they preferred using the full library of values compared to the single value system they were assigned. The full library combines multiple value systems, some of which are designed to be comprehensive in scope but differ in how individual values are defined and categorized. By combining these systems, we increase the likelihood that users can identify a value that resonates with their preferences and needs. 

Indeed, participants found that they could define the values they wanted to use more precisely with the full library:
\textit{``I like the one that has a lot of different options, because that helps me really kind of tweak what I'd be most interested in seeing (P9)''}. Some participants attributed this to the specificity of values available. For example, P10 noted that they \textit{``[were] just given more options on the full sheet and when I'm filtering out content, I don't like to filter a whole lot, because then it removes things that I might want to see, and I'd rather just scroll through [the full library] and pick it out on my own.''}

Participants also appreciated the various ways in which different systems conceptualized the values. Individuals had varying levels of understanding of values and found some values to be \textit{``kind of ambiguous (P11)''}, which sometimes limited their ability to use the single system. However, with the full library, users were able to find values that aligned more closely with their interpretations: \textit{``[Some additional values in the full system] made more sense to me, or seemed more concrete. (P3)''}

\section{Quantitative User Study}
\label{sec:eval}

Our qualitative study highlights how users benefit from real-time feed reranking and that users prefer the full library of values. To understand the patterns of library \textit{use} at scale, we conducted a between-subjects experiment with 257 participants who were randomly assigned access to either a single value system or the full library condition.

\subsection{Method}
We recruited participants via Prolific. To qualify, participants had to reside in the United States, be 18 years of age or older, be an active X/Twitter user, and pass quality checks (i.e., $\geq$ 95\% approval rate for 200+ tasks). Participants were compensated \$5.00 for completing the task, which was prorated based on an hourly wage of \$15.00. In the experiment, participants were randomly assigned access to the \textbf{Full} condition, where they had access to the entire library of values in Alexandria (N=44 participants) or the \textbf{Single} condition, where they only had access to one of the six singular value systems listed as follows: RecSys (N=37), Weibo (N=42), Reddit (N=32), Hofstede (N=33), Rokeach (N=33), or Maslow (N=36). 

Participants in each condition completed an onboarding process that involved installing our browser extension. Participants who had access to the full library completed the three-page extension onboarding process before gaining access to the remaining values. For those given a singular value system, they were directly shown the list of values, which ranged in size from six values (Maslow's hierarchy of needs) to 32 (Rokeach's value survey). After the feed was first reranked, participants were able to iteratively modify selected values and interact with their feed for ten minutes. Finally, participants were asked to respond to the following open-ended question: ``What values, if any, did you wish existed that you were not able to find?''

\subsection{Results}
We now explore participants' usage patterns with the extension, including what values they selected and the differences between the Full and Single conditions. On average, participants re-ranked their feeds $6.8\pm4.4$ times --- $6.5\pm2.6$ times in the Full condition and $6.9\pm4.6$ in the Single condition. In the Full condition, participants selected $48.9\% \pm 35.3$ of values, while in the Single condition, they selected $34.5\% \pm 27.8$ of the available values. Note that selecting a value can mean either up-ranking or down-ranking it in their feed.

\begin{figure*}
    \centering
    \includegraphics[width=\linewidth]{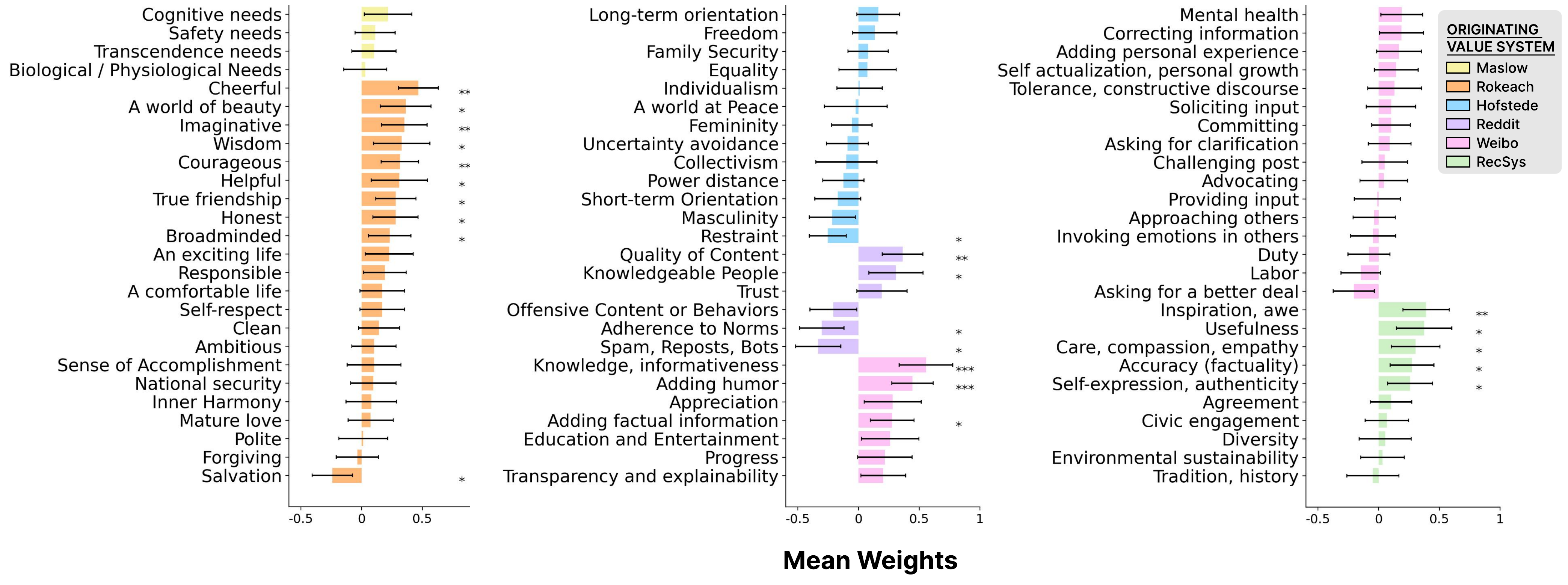}
    \caption{Twenty-three of the 78 values in the full library are significantly different ($p < 0.05$) from zero. We visualize the mean weights, which can range from -1 to 1, applied to each value by participants who had access to the full library of values. For each value, we perform a one-sample t-test with a hypothetical mean of 0. We then apply the Benjamini-Hochberg procedure on the $p$-values. Error bars represent 95\% confidence intervals. $^* p<0.05$. $^{**} p<0.01$. $^{***} p<0.001$.}
    \label{fig:meanw}
\end{figure*}

\xhdrNoPeriod{What values do participants select?}
In the Full condition, the most commonly selected values, excluding the five shown during onboarding, include ``Wisdom (75.0\%), ``Usefulness'' (75.0\%), and ``Helpful'' (72.7\%). Of the selected values, those that were often up-ranked (i.e., participants wanted more of this value in their feed) included ``Knowledge, informativeness'' (72.3\%), ``Usefulness'' (56.8\%), and ``Wisdom'' (54.5\%); values that were often downranked included ``Spam, Reposts, Bots'' (43.2\%), ``Collectivism'' (43.2\%), and ``Adherence to norms'' (40.9\%). See Appendix for the full list of value usage.

To further analyze value selections, we calculate the mean weight placed on each value across participants. We perform a one-sample t-test to see whether the mean weights are significantly different from 0, correcting the p-value using the Benjamini–Hochberg procedure to adjust for false discovery rates. This allows us to identify which values, on average, participants think should be amplified or dampened in their feed. Of the 78 values in the Full condition, $29.5\%$ (N=23) have weights significantly different from zero (Fig.~\ref{fig:meanw}). For the Single condition, the percentage of significant weights are as follows: RecSys (53.8\%, N=7), Weibo (0.0\%, N=0), Reddit (87.5\%, N=7), Hofstede (64.3\%, N=9), Rokeach (65.6\%, N=21), or Maslow (25.0\%, N=2). 

From this analysis, we identified 16 popular values, or those with mean weights significantly different from 0 in both conditions (Table~\ref{tab:popular}). Examples of values that are commonly upranked include ``Cheerful'' ($0.47$), ``Quality of Content'' ($0.36$), and ``Imaginative'' ($0.35$). Ones that are commonly downranked are ``Spam, Reposts, Bots'' ($-0.33$) and ``Restraint'' ($-0.25$). These results reflect a general consensus of what people do (or do not) want to see in their feed.

\begin{table}[htb]
    \centering
    \footnotesize
    \begin{tabular}{lrr}
    Value & Full Weight & Single Weight \\
\hline
Cheerful $\uparrow$& $0.47\pm0.16$ & $0.38\pm0.28$\\
Usefulness $\uparrow$& $0.38\pm0.23$ & $0.45\pm0.26$\\
A world of beauty $\uparrow$& $0.36\pm0.21$ & $0.35\pm0.28$\\
Quality of Content $\uparrow$& $0.36\pm0.17$ & $0.68\pm0.21$\\
Imaginative $\uparrow$& $0.35\pm0.19$ & $0.63\pm0.22$\\
Wisdom $\uparrow$& $0.33\pm0.23$ & $0.68\pm0.22$\\
Helpful $\uparrow$& $0.31\pm0.23$ & $0.44\pm0.26$\\
Knowledgeable People $\uparrow$& $0.31\pm0.22$ & $0.65\pm0.23$\\
Care, compassion, empathy $\uparrow$& $0.30\pm0.20$ & $0.41\pm0.26$\\
True friendship $\uparrow$& $0.28\pm0.17$ & $0.41\pm0.26$\\
Honest $\uparrow$& $0.28\pm0.19$ & $0.51\pm0.25$\\
Accuracy (factuality) $\uparrow$& $0.28\pm0.18$ & $0.49\pm0.26$\\
Self-expression, authenticity $\uparrow$& $0.26\pm0.19$ & $0.56\pm0.24$\\
Broadminded $\uparrow$& $0.23\pm0.17$ & $0.47\pm0.27$\\
Restraint $\downarrow$& $-0.25\pm0.15$ & $-0.45\pm0.25$\\
Spam, Reposts, Bots $\downarrow$& $-0.33\pm0.19$ & $-0.70\pm0.19$\\
    \end{tabular}
    \caption{The 16 values with significant usage (t-test $p_{adj}<0.05$) across participants in both conditions (i.e., they access to either a single value system or the full library of values).
    }
    \label{tab:popular}
\end{table}

\xhdr{The full library allows users to articulate preferences more precisely}
While we have identified values commonly used across conditions, a natural question arises: how does value selection differ between users with access to a single value system and those with access to the full library? To identify value weight differences between the two groups, we conducted a two-sample independent t-test between the weights assigned to a value across conditions. In total, we identify 18 values for which the weights are significantly different (Table~\ref{tab:displacement}). For all but one of the values, we observe that the value is \textit{displaced}, meaning that the absolute weight placed on the value is lower in the full library condition or is indistinguishable from zero. In other words, participants with access to only one value system tend to use that value, whereas participants with access to the full library--despite having access to it--give it less attention, favoring other values instead. One explanation for why value displacement occurs is that, when given access to the full library, participants can be more precise about the values they want to configure. Participants can specify how their feed must be ranked by either using a replacement value or a combination of other values. We observe weights for ``Equality'' significantly decrease from $0.50\pm0.26$ to $0.07\pm0.24$ ($p=0.02$). It was possibly replaced by more specific values, such as ``Broadminded'' ($0.23$), ``Care, compassion, empathy'' ($0.30$), or ``Tolerance'' ($0.13$). The only outlier to this pattern is ``Adherence to Norms'', which is, on average, ignored when presented in its single value system, but downranked in the full library ($0.06\pm0.29$ to $-0.30\pm0.18$).

\xhdr{The full library accommodates diverse preferences} Another benefit of having an extensive library is its ability to capture the wide range of values that people want to use for re-ranking. For each value shown in the Full condition, we calculate the percentage of participants who configured this value (either for upranking or downranking). The usage rate of values ranges from $88.6\%$ for ``Knowledge, Informativeness'' to $31.8\%$ for ``Committing'', indicating that even the least frequently used values received significant attention. This suggests that a large library is necessary to accommodate the diverse preferences of users. $61.5\%$ of values are selected by less than half of the participants. Although such values may be used less often, they can be extremely important to an individual or subset of users. For example, the value ``Restraint'' was only selected by $36.3\%$ of participants but with a significant negative weight, indicating that they do not want to see it in their feed.

\xhdr{The full library addresses specific gaps in single value systems}
We observe that the full library encompasses values that participants in the Single condition expressed interest in using to re-rank their feed. From the qualitative responses provided in the post-task questionnaire, where we ask participants which values they wish were offered that they did not find, we find participants are surfacing gaps in the singular value system that would be covered with the full library of values. To analyze these responses, we reviewed the $84$ responses that expressed values the participant thought were missing from the singular value system. Two members of the research team independently coded whether the missing value is included in our full library (Cohen's $\kappa = 0.49$). We use the intersection of values that both researchers coded as being present in the full library. In total, 55 of the desired values ($65.5\%$) had counterparts in the full library. For example, one participant wanted to see ``\emph{progressive values, human rights values, art and literature values}.'' In our library, they could select from multiple options, such as ``Equality'', ``Tolerance'', ``A World of Beauty'', and ``Imaginative.'' Values that participants wanted to see but were not covered by our library included those related to material wealth (e.g., ``\emph{great wealth accumulation}'') or more specific identity-related values (e.g., ``\emph{queerness}'', ``\emph{ally-ship}'', ``\emph{cultural values}'').

\section{Discussion}

The library of values provides a lever for users to incorporate pluralistic values in their feed ranking, as well as a mechanism to allow the broader user community to contribute values that would help shape feed ranking.

\subsection{Design Implications}

\xhdr{Toward a value-based feed ranking} Alexandria demonstrates that users seek more control over their feeds and express more satisfaction in having access to a larger library.
The values we operationalize are drawn from several established frameworks in the literature, but we make no claims that the current library of values is complete. Furthermore, we acknowledge that individuals and communities may have unique values shaped by their own experiences, which may not be fully captured in the current system. Alexandria could be enhanced to support a diversity of values by enabling the end user to add values they deem important. This approach would enable the creation of a ``\textit{marketplace of values}," where users and communities propose values, and others can subscribe to and configure these values for their feed ranking. This modular organization would empower both individuals and communities in shaping social media feeds.
For example, a specific community, such as a subreddit, could create the algorithmic feed that best represents their community-level priorities, while individuals could use Alexandria to customize their feeds. Although different in its implementation, this approach can apply both to centralized platforms---through web extensions like in our implementation---and decentralized platforms, such as Bluesky---using AT protocol~\cite{kleppmann2024bluesky}.

\xhdr{Governance} To enable the value library to expand, governance is a critical issue. Should the value library be maintained by one group, similar to steering committees that guide open-source projects, empowering this group to review the accuracy and appropriateness of submitted values? Or should it be modeled after open-source ``liberal contribution'' models, where any and all values that do not violate a set of rules should be added? What procedures should determine when and whether to merge values? Who should serve as moderators, what should the criteria be, and can they be deemed ``universal''? Future work must investigate these questions to establish robust governance frameworks.

\xhdr{Balancing values and topics} 
Our library is agnostic to whether the values represent higher-level abstract constructs (e.g., ``Freedom'', ``Equality'') or more content-oriented concepts (e.g., ``Adding humor'', ``Offensive Content or Behavior''). So long as the construct has a concrete definition that can be used for labeling, we can incorporate it into the library. The benefit this provides is that users may want to choose from a mix of broader and lower-level modules when re-ranking their feed. For example, a user may be interested in seeing content related to social change, so they uprank ``Progress.'' However, if they want their feed to be more light-hearted and humorous, they may choose more content-focused modules around adding humor and cheerfulness to uprank as well. Future work should investigate the trade-offs between higher-level and lower-level values and how designers can navigate conflicting values.
Additionally, while this blend of constructs can be helpful for users, designers may want to more explicitly delineate the boundary between ``values'' and ``topics.'' Otherwise, modules in the library may be no different than filters for specific topics, such as sports or pop culture. While content-related filters may still be useful, the purpose of the extension may creep away from being focused on these more generalizable values, which transcend topics, to very specific topics or types of content.

\xhdr{Trade off between autonomy and societal outcomes}
The significant advantages of offering users more autonomy in choosing what content to prioritize in their feed come with some challenges. This system must be designed thoughtfully before deploying it to a large social network. 
Without careful design, the system may lead to a values-based filter bubble, where social media users amplify only values aligned with their ideology, reducing the diversity of content they are exposed to when reading their feed. Although increasing autonomy can help break the filter bubble created by engagement-based algorithms, it presents risks. Given the correlation between values and political orientations \cite{Smith2017Intuitive,graham2013moral}, such as ``Conservation'' for conservatives and ``Universalism'' for liberals, promoting specific values could reinforce existing biases and beliefs. These issues already emerge with the current approaches, but without giving users explicit controls to leave the bubble. Platform designers must address this challenge by developing mechanisms and nudges that encourage exposure to diverse content, helping prevent the isolation of perspectives. In the long term, greater control over the content consumed could influence public discourse and media consumption, contributing to either more cohesive or more polarized social environments. Therefore, platforms should consider the broader societal impact of their design choices and continuously evaluate their outcomes.

\subsection{Limitations}
The current implementation of Alexandria has several limitations. First, we do not have access to the full platform inventory, so the extension can only re-rank content that is already present in the user’s feed. Second, our implementation focuses on post-level classification on Twitter/X, with classifiers that handle individual posts only. Future work should expand the library to support additional social media platforms and feed-level values, which would require access to the full set of candidate posts considered by the algorithm. Different platforms may promote different types of content and values---depending on factors such as whether content is pulled from a user’s network or from a broader pool, or what kinds of community norms are encouraged. As a result, value-based reranking might surface different kinds of content depending on the platform. Exploring these differences is an important direction for future work. Third, the classifiers themselves have limitations, as zero-shot LLM classifiers may perform inconsistently across different values. The current system uses API calls to third-party LLMs, which introduces potential data privacy risks. Although the majority of tweets that we process are already public, future iterations of Alexandria can also be run using a local instance of an LLM, eliminating the need to send any data to external servers.
Finally, the design of the tool is important, and choices such as which values are presented during onboarding may influence, at least in a short-term experiment, which values users decide to explore.

\subsection{Future Work}
One key area of future work is the long-term impact of increasing user agency.
Does using Alexandria yield any long-term attitudinal or behavioral changes? The changes executed by an extension such as Alexandria have the potential to directly impact the visibility of polarizing content. A longitudinal evaluation would also allow us to understand what values people configure over time. Do people revert to engagement ranking, confirming platforms' argument that engagement is indeed preferred by their users? Does a value-based ranking reduce engagement? Are users happy with these trade-offs?

Our approach currently only operates on the web client for social media platforms. Mobile applications of centralized social media platforms are generally restricted from client modification of this sort. This restriction limits our ability to extend our extension to the mobile experience. Other more open platforms, such as Mastodon and Bluesky, are much more obvious candidates if mobile interventions are desired since feed ranking can be performed with server-side customization. Future work should investigate how Alexandria customization can be extended to mobile devices. Finally, future work must investigate the governance structures to support a marketplace of value and what moderation approaches can be implemented. 

\section{Conclusion}
Rather than having values be tacitly embedded into our social media feeds, we present a design that gives users the ability to configure their feeds based on the values they want to see. Instead of taking a prescriptive approach and selecting one prevailing value system, we present a \emph{library} of values that can be used for re-ranking social media feeds. We instantiate our library in a Chrome browser extension, Alexandria, that empowers users to articulate their own value configurations and re-order their X/Twitter feed in real-time. Through an online study, we validate that having access to this full library of value affords users greater flexibility and precision to express the values they want to see in comparison to a single value system approach. Alexandria articulates a new vision for platform designs that empower users by offering more control over what they want to see.

\section{Ethical Considerations} 
The current implementation relies on OpenAI's API, which may cause privacy concerns for users who have not explicitly consented to sharing their feeds' content or values with a third-party service. Future developments should explore running dedicated local LLM instances to address these concerns. Participant usage data logs (including value rankings and Twitter feeds) are stored in a database hosted on a secure cloud instance. Data is only accessible to members of the research team. After installing the extension, the user's Twitter ID is hashed locally before any data is sent to our database, meaning we only store deidentified data. Recordings of the qualitative interviews were transcribed using Otter.ai and will be deleted in 3 months. This work has been approved by the IRB of our institution to ensure adherence to ethical standards.

\section{Acknowledgments}
We thank members of the Stanford HCI Group and Social Media AI Reading Group for their helpful feedback. DZ acknowledges support from the Brown
Institute for Media Innovation. This work was sponsored by the Hoffman-Yee Research Grants at the Stanford Institute for Human-Centered Artificial Intelligence (HAI).

\bibliography{references}

\subsection{Paper Checklist}

\begin{enumerate}

\item For most authors...
\begin{enumerate}
    \item  Would answering this research question advance science without violating social contracts, such as violating privacy norms, perpetuating unfair profiling, exacerbating the socio-economic divide, or implying disrespect to societies or cultures?
    \answerYes{Yes.}
  \item Do your main claims in the abstract and introduction accurately reflect the paper's contributions and scope?
    \answerYes{Yes.}
   \item Do you clarify how the proposed methodological approach is appropriate for the claims made? 
    \answerYes{Yes. See introduction of the 2 studies.}
   \item Do you clarify what are possible artifacts in the data used, given population-specific distributions?
    \answerNA{NA}
  \item Did you describe the limitations of your work?
    \answerYes{See Sec. Limitations.}
  \item Did you discuss any potential negative societal impacts of your work?
    \answerYes{Yes, see Sec. Ethical Considerations.}
      \item Did you discuss any potential misuse of your work?
    \answerYes{Yes, see Sec. Ethical Considerations.}
    \item Did you describe steps taken to prevent or mitigate potential negative outcomes of the research, such as data and model documentation, data anonymization, responsible release, access control, and the reproducibility of findings?
    \answerYes{Yes. See the Method subsection of the two studies.}
  \item Have you read the ethics review guidelines and ensured that your paper conforms to them?
    \answerYes{Yes.}
\end{enumerate}

\item Additionally, if your study involves hypotheses testing...
\begin{enumerate}
  \item Did you clearly state the assumptions underlying all theoretical results?
    \answerNA{NA}
  \item Have you provided justifications for all theoretical results?
    \answerNA{NA}
  \item Did you discuss competing hypotheses or theories that might challenge or complement your theoretical results?
    \answerNA{NA}
  \item Have you considered alternative mechanisms or explanations that might account for the same outcomes observed in your study?
    \answerNA{NA}
  \item Did you address potential biases or limitations in your theoretical framework?
    \answerNA{NA}
  \item Have you related your theoretical results to the existing literature in social science?
    \answerNA{NA}
  \item Did you discuss the implications of your theoretical results for policy, practice, or further research in the social science domain?
    \answerNA{NA}
\end{enumerate}

\item Additionally, if you are including theoretical proofs...
\begin{enumerate}
  \item Did you state the full set of assumptions of all theoretical results?
    \answerNA{NA}
	\item Did you include complete proofs of all theoretical results?
    \answerNA{NA}
\end{enumerate}

\item Additionally, if you ran machine learning experiments...
\begin{enumerate}
  \item Did you include the code, data, and instructions needed to reproduce the main experimental results (either in the supplemental material or as a URL)?
    \answerYes{Yes, see Appendix for model prompts.}
  \item Did you specify all the training details (e.g., data splits, hyperparameters, how they were chosen)?
    \answerNA{We do not train our own machine learning model. We use \texttt{GPT-4o-mini}.}
     \item Did you report error bars (e.g., with respect to the random seed after running experiments multiple times)?
    \answerYes{Yes.}
	\item Did you include the total amount of compute and the type of resources used (e.g., type of GPUs, internal cluster, or cloud provider)?
    \answerYes{Yes, when describing the system, we discuss that we use \texttt{GPT-4o-mini} via the OpenAI API.}
     \item Do you justify how the proposed evaluation is sufficient and appropriate to the claims made? 
    \answerYes{Yes, see Appendix.}
     \item Do you discuss what is ``the cost`` of misclassification and fault (in)tolerance?
    \answerNA{NA}
  
\end{enumerate}

\item Additionally, if you are using existing assets (e.g., code, data, models) or curating/releasing new assets, \textbf{without compromising anonymity}...
\begin{enumerate}
  \item If your work uses existing assets, did you cite the creators?
    \answerNA{NA}
  \item Did you mention the license of the assets?
    \answerNA{NA}
  \item Did you include any new assets in the supplemental material or as a URL?
    \answerNA{NA}
  \item Did you discuss whether and how consent was obtained from people whose data you're using/curating?
    \answerNA{NA}
  \item Did you discuss whether the data you are using/curating contains personally identifiable information or offensive content?
    \answerNA{NA}
\item If you are curating or releasing new datasets, did you discuss how you intend to make your datasets FAIR?
\answerNA{NA}
\item If you are curating or releasing new datasets, did you create a Datasheet for the Dataset? 
\answerNA{NA}
\end{enumerate}

\item Additionally, if you used crowdsourcing or conducted research with human subjects, \textbf{without compromising anonymity}...
\begin{enumerate}
  \item Did you include the full text of instructions given to participants and screenshots?
    \answerYes{Yes, see Appendix.}
  \item Did you describe any potential participant risks, with mentions of Institutional Review Board (IRB) approvals?
    \answerYes{Yes.}
  \item Did you include the estimated hourly wage paid to participants and the total amount spent on participant compensation?
    \answerYes{Yes, see the Qualitative User Study and Quantitative User Study.}
   \item Did you discuss how data is stored, shared, and deidentified?
   \answerYes{Yes. see Sec. Ethical Considerations}
\end{enumerate}

\end{enumerate}

\clearpage

\appendix
\renewcommand{\thefigure}{A.\arabic{figure}}  
\renewcommand{\thetable}{A.\arabic{table}}    
\renewcommand{\thelstlisting}{A.\arabic{lstlisting}}  

\setcounter{figure}{0}   
\setcounter{table}{0}    
\setcounter{lstlisting}{0}  

\section{Appendix}

\subsection{Framework for Library Value Sourcing}

\begin{figure}[tbh]
  \centering
  \includegraphics[width=\linewidth]{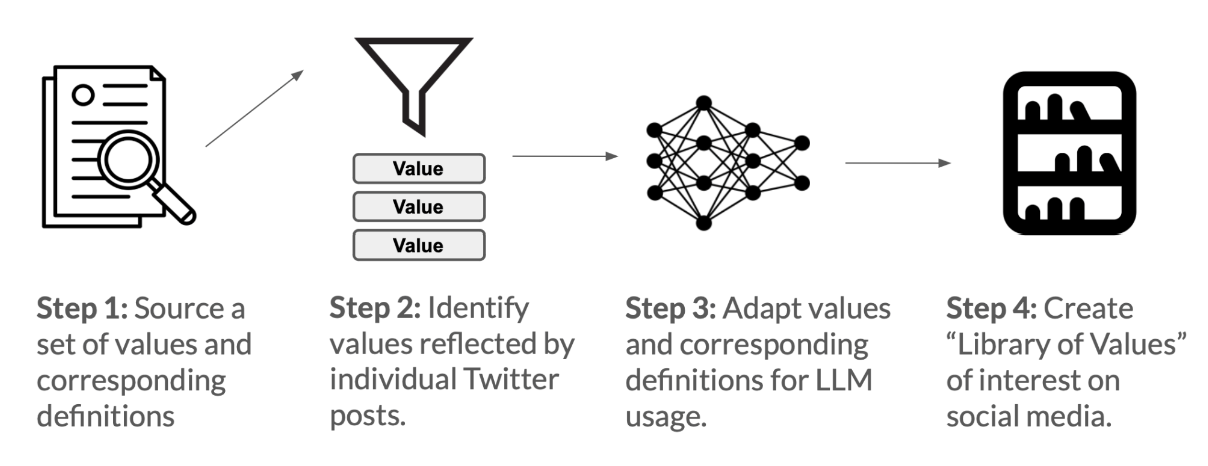}
  \caption{Our value library begins with a set of values for social media algorithms drawn from existing literature and frameworks. We adapt these frameworks' terms and definitions, then use a large language model (LLM) pipeline to annotate expressions of those values in social media content, creating algorithmic operationalizations of each value.}
  \label{fig:library}
\end{figure}

\subsection{LLM Prompt For Labeling Values}

We used the LLM prompt in Listing A1 to label values in individual tweets. Each tweet was labeled separately using GPT-4o-mini, with both the tweet's text content and any associated images (in PNG format) included in the prompt. All values that user's assigned a non-zero weight to were evaluated simultaneously.

\begin{lstlisting}[style=prompt, caption={The prompt used for labeling social media content. We replace \texttt{\$conceptDefinitions} with a dictionary of values that users selected to uprank or downrank and their corresponding definitions. See the Appendix for the definitions of all the translated values.}, label={lst:prompt}]
Consider the following set of concepts, listed as CONCEPT : DEFINITION

${conceptDefinitions}

For each concept, rate whether the tweet reflects the concept using the following scale:
  0 = This post does not reflect this concept, or it is unclear
  1 = This post weakly reflects this concept
  2 = This post strongly reflects this concept
  
Output: One JSON dictionary of the following format. Rating is a dictionary of key value pairs, with each key being a concept and each value your rating for that concept: {"Rating": {"Concept": RATING} }
 
Do not output anything else. If you don't know, make your best guess. Do not ask for an example.
\end{lstlisting}

\subsection{Evaluation of the Value Classifier}
\label{appendix-llm-evaluation}
Our algorithmic value annotation method relies on the ability of large language models to accurately label content, given a sufficiently precise definition from the literature~\cite{bernstein2023embedding}. Prior work has demonstrated that this approach agrees with experts as much as experts agree with each other on social media posts (e.g.,~\cite{jia2023embedding}). Nevertheless, we ran an annotation task to ascertain whether the classification of the saliency of values in our library is aligned with human's perception of them. 

\subsubsection{Method}
We sampled 2 values randomly from each value system for this evaluation, which resulted in 12 values in total. We sourced the tweets for annotation from the Quantitative User Study by selecting a random sample of 10,000 posts from real feeds of users. We used an LLM prompt to filter our not-safe-for-work content and, given that certain tweets lack complete context needed for comprehension by an independent rater, we also filtered out incomprehensible posts. 

\begin{lstlisting}[style=prompt, caption={The prompt used for labeling comprehensibility of tweets. We only kept tweets with a label of 3.}, label={lst:prompt}]
Task -- I will provide a Twitter post alongside a codebook that describes in which settings Twitter posts are comprehensible Tweets, and in which settings Tweets lack understandability. I want you to apply each concept in the codebook to determine why and to what degree the concept the applies to the post. As you answer, please take the following steps:
Step 1) For each concept in the codebook, describe whether and to what degree the Tweet illustrates comprehensible, or uncomprehensible behavior, with the following format: {"<CONCEPT 1>": {"Why": "<explanation of how the concept applies to the post>", "Rating": <rating>}, ..., "<CONCEPT N>": {"Why": "<explanation of how the concept applies to the post>", "Rating": <rating>}}. ("Codebook Application")
Use the following scale to assign your rating:
    0="the post strongly exhibits uncomprehensible behavior for the given concept"
    1="the post somewhat exhibits uncomprehensible behavior for the given concept"
    2="the post somewhat exhibits comprehensible behavior for the given concept"
    3="the post strongly exhibits comprehensible behavior for the given concept"
Step 2) Summarizing your reasoning in Steps 1 and 2, determine a single rating for whether the post is comprehensible, or uncomprehensible: "Final Rating": {"Why": "<explanation of final rating>", "Rating": <rating>}. ("Agreement Rating")
Use the following scale to assign your rating:
    0="the post strongly exhibits uncomprehensible behavior"
    1="the post somewhat exhibits uncomprehensible behavior"
    2="the post somewhat exhibits comprehensible behavior"
    3="the post strongly exhibits comprehensible behavior"

Codebook --
READABILITY: READABLE -- Well-structured, coherent, and easily understandable language that facilitates the comprehension of the expressed values; UNREADABLE -- Poorly structured, incoherent, or overly complex language that hinders understanding of the message
COHERENCE: COHERENT -- Logically structured, with a clear progression of ideas that underscore the expressed meaning; INCOHERENT -- Disjointed or lacking logical flow, making it difficult to discern any underlying meaning
SPAM BEHAVIOR: AUTHENTIC -- Content is original, personal, and specifically targets relevant issues; SPAMMY -- Promotional, repetitive, or unsolicited content or appears automated
CONTEXT REQUIRED FOR UNDERSTANDING: SELFCONTAINED -- Minimal to no additional context required, as the post is self-contained and clear; REQUIRESCONTEXT -- Requires external information or additional context for understanding due to vagueness or lack of focus


Output Format -- Output your message in JSON, where you provide the following:
{
    "Codebook Application": {
        "<CONCEPT 1>": {
            "Why": "<explanation of how the concept applies to the post>", "Rating": <integer from 0-3>}, 
        ..., 
        "<CONCEPT N>": {
            "Why": "<explanation of how the concept applies to the post>", "Rating": <integer from 0-3>}
    },
    "Final Rating": {
            "Why": "<explanation of final rating>", "Rating": <integer from 0-3>}
}
\end{lstlisting}

\begin{lstlisting}[style=prompt, caption={The prompt use to identify NSFW content. Only tweets with an overall rating of 0 were kept}, label={lst:promptNSFW}]
Task -- I will provide a Twitter post alongside a codebook that describes in which settings Twitter posts are NSFW Tweets (not safe for work). I want you to apply each concept in the codebook to determine why and to what degree the concept the applies to the post. As you answer, please take the following steps:
Step 1) For each concept in the codebook, describe whether and to what degree the Tweet illustrates NSFW behavior, with the following format: {"<CONCEPT 1>": {"Why": "<explanation of how the concept applies to the post>", "Rating": <rating>}, ..., "<CONCEPT N>": {"Why": "<explanation of how the concept applies to the post>", "Rating": <rating>}}. ("Codebook Application")
Use the following scale to assign your rating:
    0="This post does not contain NSFW content at all"
    1="This post contains a little bit of NSFW content"
    2="This post moderately contains NSFW content"
    3="This post strongly contains NSFW content"
Step 2) Summarizing your reasoning in Steps 1 and 2, determine a single rating for whether the post is NSFW: "Final Rating": {"Why": "<explanation of final rating>", "Rating": <rating>}. ("Agreement Rating")
Use the following scale to assign your rating:
    0="This post does not contain NSFW content at all"
    1="This post contains a little bit of NSFW content"
    2="This post moderately contains NSFW content"
    3="This post strongly contains NSFW content"

Codebook --
SEXUAL: SEXUAL -- Contains graphic imagery or text that is sexually explicit; NOTSEXUAL -- Does not contain sexually explicit imagery. Safe for the workplace.
VIOLENT: VIOLENT -- Contains graphic imagery or text that describes violence in an explicit or egregieous way; NONVIOLNET -- Contains no reference to violence
DEROGATORY: DEROGATORY -- Contains derogatory imagery or text that could make someone uncomfortable; NONDEROGATORY -- Does not contain any derogatory imagery or text


Output Format -- Output your message in JSON, where you provide the following:
{
    "Codebook Application": {
        "<CONCEPT 1>": {
            "Why": "<explanation of how the concept applies to the post>", "Rating": <integer from 0-3>}, 
        ..., 
        "<CONCEPT N>": {
            "Why": "<explanation of how the concept applies to the post>", "Rating": <integer from 0-3>}
    },
    "Final Rating": {
            "Why": "<explanation of final rating>", "Rating": <integer from 0-3>}
}
\end{lstlisting}

For each value, we selected a stratified sample of 30 tweets, evenly distributed across the LLM's classifications: not reflecting, weakly reflecting, and strongly reflecting the value. This approach increased the chances of including positive examples, as some values are rare on social media and ground truth labels were not yet available. The tweets for each value were added to a Qualtrics survey that asked the participant to rate the extent to which each tweet reflects the given concept on the same Likert scale as the classifier. The survey also provided the definition of the value. 

We recruited participants for the task via Prolific. To qualify, participants had to reside in the United States, be 18 years of age or older, and pass quality checks (i.e., $\geq$ 95\% approval rate for 200+ tasks). For each value, a total of 5 participants completed the annotation task. Each participant was compensated \$3 for completing the task.

\subsubsection{Results}
To obtain a human score for the value saliency of each tweet, we averaged the labels by the 5 annotators and then rounded the average to the nearest integer. We then compared this human score for the value with the score given to the tweet by the LLM. The mappings from labels to numbers were the following: Not reflecting the value to 0, weakly reflecting to 1, and strongly reflecting to 2.

The extent to which the classifier and average human labels agreed on the presence of absence of tweets regardless of how saliently a value was perceived as present (binary classification) ranged from 0.69 (for the value ``Restraint'') to 1 (for the value ``Biological and Physiological needs''), as shown in Table~\ref{tab:metrics_comparison}. We also report the binary F1 score and recall in Table~\ref{tab:metrics_comparison}.

We also calculated the Mean Absolute Error (MAE) between the predicted and the ground truth labels to evaluate the extent of disagreement, accounting for both the presence of a value and the magnitude of misalignment. Unlike binary classification, which only considers agreement on presence or absence, MAE reflects how far the predicted label deviates from the human-annotated saliency of the value. MAE ranged from 0.28 (for value biological and physiological needs) to 0.75 (for value restraint). 

We compare the performance of our classifier to the performance of an individual rater in estimating the consensus (the average rating). Using the five human votes on each tweet, we calculate the average human vote mean absolute error (HMAE) for each of the values, by taking the absolute error between a human vote and the average of the four other votes. 

For 10 of the 12 values, we find the classifier is a better predictor of the average human votes than the individual humans are---on average, our classifier performs worse only for values restraint and diversity. This indicates for most values, our classifier performs as well as, if not better than, individuals for estimating the mean. Indeed, on average, the classifier MAE is 0.161 lower than the HMAE.

\begin{table*}[h]
\label{tab:classifier-eval}
\centering
\footnotesize
\begin{tabular}{lcccccc}
\toprule
\textbf{Value} & \textbf{Binary \%} & \textbf{F1} & \textbf{Recall} & \textbf{Precision} & \textbf{MAE} & \textbf{HMAE} \\ 
\midrule
Bio. and Physiological Needs & 100.0 & 1.000 & 1.000 & 1.000 & 0.281 & 0.568 \\
Self-respect & 90.0 & 0.919 & 0.895 & 0.944 & 0.333 & 0.568 \\
Transcendence needs & 88.2 & 0.913 & 0.875 & 0.955 & 0.441 & 0.594 \\
National-security & 87.5 & 0.900 & 0.818 & 1.000 & 0.313 & 0.578 \\
Care, compassion, empathy & 87.1 & 0.895 & 0.810 & 1.000 & 0.355 & 0.452 \\
Diversity & 80.0 & 0.833 & 0.750 & 0.938 & 0.400 & 0.375 \\
Long-term orientation & 76.7 & 0.837 & 0.900 & 0.783 & 0.567 & 0.742 \\
Invoking emotions in others & 75.0 & 0.846 & 1.000 & 0.733 & 0.469 & 0.732 \\
Spam, Reposts, Bots & 74.2 & 0.846 & 1.000 & 0.733 & 0.613 & 0.786 \\
Knowledgeable People & 73.3 & 0.833 & 1.000 & 0.714 & 0.467 & 0.664 \\
Knowledge, informativeness & 73.3 & 0.826 & 0.950 & 0.731 & 0.400 & 0.625 \\
Restraint & 68.8 & 0.762 & 0.727 & 0.800 & 0.750 & 0.633 \\
\midrule
Average Across Values & 81.2 & 0.866 & 0.888 & 0.853 & 0.449 & 0.610 \\
\bottomrule
\end{tabular}
\color{black}
\caption{Comparison of LLM Predictions vs Human Annotations. MAE is the mean absolute error of the rounded human average and actual votes. Binary \% is the percent accuracy on the binary value classification task between the rounded average human raters and the LLM. HMAE is the mean absolute error of a human vote compared to the average of the other four votes.}
\label{tab:metrics_comparison}
\end{table*}

\subsection{Evaluating the Recognizability of Value-Aligned Feeds using LLM-as-a-Judge}
\label{appendix:llm-as-a-judge}

The end goal of our browser extension and ranking method is to produce feeds that are \textit{more value aligned} and therefore different from engagement feeds. Our user evaluation---both quantitative and qualitative---directly measures value usage and overall satisfaction. As a supplementary analysis to further validate our pipeline, we perform experiments to demonstrate the difference between value-aligned and existing engagement feeds. Concretely, we use all value-aligned feeds created during quantitative user evaluation study and their corollary engagement feed. Then, we evaluate whether the value-aligned feed or the engagement feed is more likely to be ranked using the set of selected values.  We use an LLM-as-a-judge (\texttt{Claude-sonnet-3.7}, \texttt{GPT-4o}, \texttt{Gemini-2.5-flash-pro}) as our evaluator. Across the three models, the LLM evaluator recognized the value-aligned feed $71.3\%$ of the time, with \texttt{Gemini-2.5-flash-pro} having the highest accuracy at $77.0\%$. \texttt{GPT-4o} and \texttt{Claude-sonnet-3.7} achieved an accuracy of $76.0\%$ and $61.0\%$ respectively.

{
\begin{lstlisting}[style=prompt, caption={The prompt used for LLM of the judge to rank feeds. We replace tweets with the tweet text and descriptions of each image, generated using gpt-4o-mini} label={lst:prompt2}]
You are an expert judge of Twitter feeds. You will be given two feeds, one feed is a default feed, and one feed is ranked according to specific values in the tweets. Your job is to identify which is the value ranked feed.

The following values were used to rank the feed. A positive value indicates the value should be brought to the top of the feed (higher priority), while a negative value indicates the value should be demoted (lower priority). The absolute value of the number indicates how much the value should be prioritized or demoted. The values are as follows:\n\n

${value with weights}

Here is feed A:\n
${Tweet 1: ...}$
${Tweet 2: ...}$
...

Here is feed B:\n
${Tweet 1: ...}$
${Tweet 2: ...}$
...


Return ONLY A or B, corresponding to the feed that is ranked according to the values. Do not include any other text.}

\end{lstlisting}

\subsection{Motivating Scenario For the Extension}
We provide a motivating scenario to illustrate our goals for the system. 

Tina is an active X/Twitter user who mainly uses the platform to see entertaining content that makes her laugh. However, she noticed that much of the content at the top of her feed contains people talking about their achievements, which she finds overwhelming. Although she can use built-in features to remove individual users or certain keywords from her feed, Tina's X algorithm continues to push this content.

When Tina is given access to Alexandria, she down-ranks ``Achievement'' and ``Inspiration, Awe.'' She reranks her feed and notices that posts with people bragging about their accomplishments have moved to the bottom of her feed. However, Tina still wants humorous and light-hearted content to be the first thing she sees. In order to modify her feed further, Tina selects the values ``Adding humor'' and ``Cheerful'' to her set of upranked values. After re-ranking her feed again, Tina does see more funny content at the top of her feed, but wants to make a few more modifications. She decides to use the slider toggle for ``Cheerful'' to see how her feed changes when she decreases the weight placed on this value. Tina decides that she likes a mix of happy and dark humorous content, so she decides to set the slider at a weight of $0.5$. After using Alexandria for a week, Tina is satisfied with the choices she made prior, but now she also wants to see more informative content on her feed. As a result, she upranks the values of ``Education and Entertainment'' and ``Knowledgeable People.''

\subsection{Coverage of Tweets by our Library}\label{appendix:coverage}

To highlight the practical coverage of our library, Fig.~\ref{fig:app_coverage} shows the distribution of values across the posts we labeled during the merging process and the percentage of values they contain--- less than $3\%$ of all posts contain no values, suggesting that our library can effectively rank nearly any post on X/Twitter.

\begin{figure}[hb]
    \centering
    \includegraphics[width=0.8\linewidth]{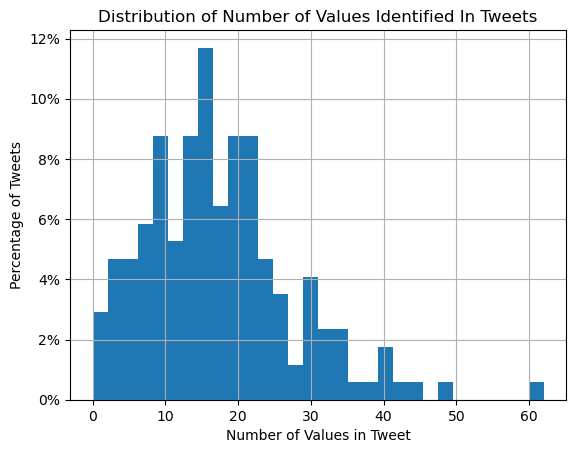}
    \caption{Distribution of number of values from our library found on Twitter posts. Less than 3\% of posts have no values present.}
    \label{fig:app_coverage}
\end{figure}

\subsection{Chrome Extension Implementation Details} 
Our extension is built for the Chrome browser, created using a Vue.js frontend and a Node.js server.  We store data collected from the extension, including which values users selected, the history of value selection, and the ordering of the re-ranked feed in a MySQL database. To label the values in the tweets, we use the OpenAI API and \texttt{GPT-4o-mini} model with a temperature of $0$.

\subsection{Qualitative Evaluation Interview Script} 

Interviewers followed the following script for the semi-structured user interviews. The interviewers asked follow-up questions to get more details/background on the questions listed below. Participants were asked about Social Media Habits before installing the extension. After using the extension with a single value system, they were asked the Individual Value System questions. After using the full library system, they were asked the library of value questions and the user experience questions. 

\subsubsection {Script}

\textit{Social Media Habits}:
This section helps contextualize the participants’ prior social media habits before they use the extension. 
What do you think your current social media feed prioritizes showing you? 
What do you like about it? 
What bothers you about it?
How do you exert control over your social media feed to show the content you want to see? How do you avoid what you don’t want to see?

\textit{Individual Value System}:
This section asks participants to provide their opinion about an extension that features only one value system. 
How satisfied are you with the customization options provided by this Feed Ranker?
Which modules did you pick to enable for reranking your feed? Why those? 
Which modules do you wish existed, but didn't? 

\textit{Library of Values Questions}:
This section exists to allow participants to express 1) their opinion about the entire library of values 2) how the library of values differs from an extension that features only one value system. 
How satisfied are you with the customization options provided by this Feed Ranker?
Which modules did you pick to enable for reranking your feed? Why those? 
Which modules do you wish existed, but didn't? 
Between this feed ranker and the previous feed ranker, what do you like about each approach, and what do you not like about each approach? Focus on how they do and don't allow you to prioritize the values that are important to you in your feed
Which one allows you to obtain outcomes that are important to your social media feed?
For both this feed ranker and the previous feed ranker, how successful were you in accomplishing a feed that reflected what you value and believe in?

\textit{User Experience Questions/Ending Questions}:
This section contains concluding questions that ask about the extension more broadly.
While using the extension, how easy or difficult was it for you to explore and find different values?
While using the extension, do you feel you have control over what you see on your feed? Why or why not?
Please compare and contrast your experience with the two feed ranker variations. What are your impressions of each feed ranker?

\begin{table}[h]
\label{tab:classifier-eval}
\centering
\begin{tabular}{cccc}
\toprule
\textbf{Participant} & \textbf{Sex} & \textbf{Race} & \textbf{Age} \\
\midrule

1  & Female & Asian            & 25-34 \\
2  & Male   & White            & 35-44 \\
3  & Male   & Asian            & 25-34 \\
4  & Female & White            & 25-34 \\
5  & Male   & White            & 45-54 \\
6  & Male   & Mixed            & 25-34 \\
7  & Female & White            & 35-44 \\
8  & Female & White            & 45-54 \\
9  & Male   & White            & 25-34 \\
10 & Male   & Black            & 25-34 \\
11 & Female & Asian            & 18-24 \\
12 & Female & Asian            & 18-24 \\
\hline
\end{tabular}
\caption{Demographic information of qualitative interview participants}
\label{tab:demographics}
\end{table}

\subsection{Quantitative Evaluation User Instructions} 

In the post-task questionnaire, we ask the following open-ended questions: 
\begin{itemize}
    \item What values, if any, did you wish existed that you were not able to find? 
    \item Do you have any additional feedback about your experience that you would like to provide? Did you encounter any bugs?
\end{itemize}

\begin{table*}[ht]
    \centering
    \begingroup
    \tiny 
    \begin{tabular}{p{0.30\textwidth} p{0.15\textwidth} p{0.50\textwidth}}
    \toprule

    \textbf{Value} & \textbf{Original System} & \textbf{Definition} \\ 
    \midrule
    Mature love & Rokeach & Deep emotional and physical intimacy. \\
    An exciting life & Rokeach & A life full of stimulation and activity. \\
    Salvation & Rokeach & Spiritual deliverance and eternal life. \\
    A comfortable life & Rokeach & A life marked by prosperity and comfort. \\
    Self-respect & Rokeach & Maintaining self-esteem and dignity. \\
    Inner Harmony & Rokeach & Freedom from internal conflict and stress. \\
    True friendship & Rokeach & Close and enduring companionship. \\
    Ambitious & Rokeach & Being hard-working and aspiring to success. \\
    Cheerful & Rokeach & Being lighthearted and joyful. \\
    Broadminded & Rokeach & Being open to new ideas and perspectives. \\
    Clean & Rokeach & Maintaining neatness and tidiness. \\
    A world of beauty & Rokeach & Appreciation for nature and the arts. \\
    Responsible & Rokeach & Being dependable and reliable. \\
    Wisdom & Rokeach & A mature understanding and insight into life. \\
    Imaginative & Rokeach & Being daring and creative. \\
    Polite & Rokeach & Being courteous and well-mannered. \\
    A sense of accomplishment & Rokeach & Making a lasting contribution or achievement. \\
    Honest & Rokeach & Being sincere and truthful. \\
    Forgiving & Rokeach & Being willing to pardon others. \\
    National security & Rokeach & Protection from external threats and attacks. \\
    Helpful & Rokeach & Working for the welfare of others. \\
    Courageous & Rokeach & Standing up for your beliefs. \\
    Soliciting input & Weibo & Seeking advice or additional information from others. \\
    Correcting information & Weibo & Fixing errors or inaccuracies in shared content. \\
    Providing input & Weibo & Contributing targeted information to discussions. \\
    Committing & Weibo & Promising to take action related to a post. \\
    Transparency and explainability & Weibo & Providing clear and truthful information about systems. \\
    Labor & Weibo & Engaging in meaningful work and its impact on well-being. \\
    Self actualization, personal growth & Weibo & Reaching full potential and personal growth. \\
    Tolerance, constructive discourse & Weibo & Encouraging respectful and productive discussions. \\
    Appreciation & Weibo & Expressing positive emotions or recognition. \\
    Adding personal experience & Weibo & Sharing personal accounts to enrich discussions. \\
    Adding factual information & Weibo & Providing additional accurate details or data. \\
    Approaching others & Weibo & Engaging others with small talk or greetings. \\
    Advocating & Weibo & Supporting or recommending products or places. \\
    Invoking emotions in others & Weibo & Tagging others while sharing personal feelings. \\
    Asking for a better deal & Weibo & Negotiating for discounts or better offers. \\
    Adding humor & Weibo & Incorporating humor in non-humorous contexts. \\
    Progress & Weibo & The momentum toward better technology and quality of life. \\
    Challenging post & Weibo & Questioning or objecting to content for improvement. \\
    Asking for clarification & Weibo & Requesting further explanation or information. \\
    Mental health & Weibo & Promoting psychological well-being and coping with stress. \\
    Duty & Weibo & Obligation to fulfill responsibilities beyond self-interest. \\
    Knowledge, informativeness & Weibo & Keeping users informed on relevant topics. \\
    Education and Entertainment & Weibo & Commenting on the value of content for learning or enjoyment. \\
    Cognitive needs & Maslow & The need for knowledge and understanding. \\
    Transcendence needs & Maslow & Helping others achieve self-actualization. \\
    Biological and Physiological needs & Maslow & Basic survival needs like food, shelter, and sleep. \\
    Safety needs & Maslow & Protection from harm and ensuring stability. \\
    Family Security & Hofstede & Ensuring the well-being of loved ones. \\
    Short-term Orientation & Hofstede & Focus on tradition, immediate outcomes, and steadfastness. \\
    Restraint & Hofstede & Controlling gratification of desires through strict social norms. \\
    Power distance & Hofstede & Acceptance of unequal power distribution in organizations and society. \\
    Uncertainty avoidance & Hofstede & Society's tolerance for ambiguity and unexpected events. \\
    Collectivism & Hofstede & Tight-knit societies where extended families and others form in-groups. \\
    Masculinity & Hofstede & A preference for achievement, assertiveness, and material success. \\
    Individualism & Hofstede & The degree to which individuals are integrated into groups. \\
    Freedom & Hofstede & Independence and the ability to make free choices. \\
    Femininity & Hofstede & A preference for cooperation, modesty, and quality of life. \\
    Long-term orientation & Hofstede & Society's focus on future rewards and perseverance. \\
    Equality & Hofstede & Equal opportunity and fairness for all. \\
    A world at Peace & Hofstede & A society free of war and conflict. \\
    Civic engagement & RecSys & Actively participating in community and public life. \\
    Agreement & RecSys & Expressing acceptance or concurrence with others' views. \\
    Diversity & RecSys & Appreciating and respecting cultural differences. \\
    Tradition, history & RecSys & Valuing heritage, practices, and communal identity. \\
    Self-expression, authenticity & RecSys & Expressing identity and individuality openly. \\
    Accuracy (factuality) & RecSys & Ensuring the correctness and reliability of information. \\
    Environmental sustainability & RecSys & Respecting and preserving the natural environment. \\
    Inspiration, awe & RecSys & Seeking guidance, motivation, and goals in uncertain times. \\
    Care, compassion, empathy & RecSys & Emphasizing interdependence and responsibility for others. \\
    Usefulness & RecSys & Providing relevant and helpful services or information. \\
    Adherence to Norms & Reddit & Following and respecting community rules and standards. \\
    Offensive, Abusive, Harassing Content or Behaviors & Reddit & Addressing content or behaviors that are harmful or disrespectful. \\
    Knowledgeable People & Reddit & Valuing the expertise and credibility of community members. \\
    Quality of Content & Reddit & Evaluating the usefulness and value of shared content. \\
    Spam, Reposts, Bots & Reddit & Commenting on the presence of unwanted or automated content. \\
    Trust & Reddit & Assessing the trustworthiness of people and content in communities. \\
    \bottomrule
    \end{tabular}
    \caption{Our library of values consists of 78 value constructs sourced from six popular value systems. We list all values with the original system they were sourced from and the definition used in our LLM labeler.}
    \label{tab:myxtabular}
    \endgroup
\end{table*}
\clearpage
\begin{table*}[]
    \centering
     \tiny 
    \begin{tabular}{lrrr}
    \toprule 
    \textbf{Value} & \textbf{Usage Rate (\%)} & \textbf{Upranked (\%)} & \textbf{Downranked (\%)}\\
    \midrule
    Knowledge, informativeness & 88.6 & 72.7 & 15.9\\
    A world at Peace & 79.5 & 38.6 & 40.9\\
    Collectivism & 77.3 & 34.1 & 43.2\\
    Wisdom & 75.0 & 54.5 & 20.5\\
    Usefulness & 75.0 & 56.8 & 18.2\\
    Appreciation & 72.7 & 50.0 & 22.7\\
    Helpful & 72.7 & 52.3 & 20.5\\
    Education and Entertainment & 72.7 & 50.0 & 22.7\\
    Knowledgeable People & 68.2 & 50.0 & 18.2\\
    Equality & 65.9 & 36.4 & 29.5\\
    Progress & 63.6 & 43.2 & 20.5\\
    A world of beauty & 63.6 & 50.0 & 13.6\\
    Tolerance, constructive discourse & 59.1 & 36.4 & 22.7\\
    Inspiration, awe & 59.1 & 50.0 & 9.1\\
    A sense of accomplishment & 59.1 & 34.1 & 25.0\\
    Tradition, history & 56.8 & 27.3 & 29.5\\
    Care, compassion, empathy & 56.8 & 43.2 & 13.6\\
    Diversity & 54.5 & 29.5 & 25.0\\
    Cheerful & 54.5 & 50.0 & 4.5\\
    Trust & 54.5 & 36.4 & 18.2\\
    Adding humor & 54.5 & 50.0 & 4.5\\
    Imaginative & 54.5 & 45.5 & 9.1\\
    Spam, Reposts, Bots & 52.3 & 9.1 & 43.2\\
    Adherence to Norms & 52.3 & 11.4 & 40.9\\
    Inner Harmony & 52.3 & 29.5 & 22.7\\
    Offensive, Abusive, Harrassing Content or Behaviors & 50.0 & 13.6 & 36.4\\
    Soliciting input & 50.0 & 29.5 & 20.5\\
    An exciting life & 50.0 & 36.4 & 13.6\\
    Honest & 50.0 & 38.6 & 11.4\\
    Cognitive needs & 50.0 & 36.4 & 13.6\\
    Self-expression, authenticity & 47.7 & 36.4 & 11.4\\
    Masculinity & 47.7 & 13.6 & 34.1\\
    Adding factual information & 47.7 & 38.6 & 9.1\\
    Polite & 47.7 & 25.0 & 22.7\\
    Advocating & 47.7 & 27.3 & 20.5\\
    Accuracy (factuality) & 47.7 & 38.6 & 9.1\\
    Individualism & 45.5 & 22.7 & 22.7\\
    Quality of Content & 45.5 & 40.9 & 4.5\\
    Short-term Orientation & 45.5 & 13.6 & 31.8\\
    Broadminded & 43.2 & 34.1 & 9.1\\
    Transparency and explainability & 43.2 & 31.8 & 11.4\\
    Providing input & 43.2 & 20.5 & 22.7\\
    Self-respect & 43.2 & 29.5 & 13.6\\
    A comfortable life & 43.2 & 29.5 & 13.6\\
    Correcting information & 43.2 & 31.8 & 11.4\\
    Mature love & 43.2 & 25.0 & 18.2\\
    Challenging post & 43.2 & 25.0 & 18.2\\
    Invoking emotions in others & 43.2 & 18.2 & 25.0\\
    National security & 43.2 & 27.3 & 15.9\\
    Freedom & 43.2 & 27.3 & 15.9\\
    Adding personal experience & 43.2 & 29.5 & 13.6\\
    Asking for a better deal & 40.9 & 9.1 & 31.8\\
    Transcendence needs & 40.9 & 25.0 & 15.9\\
    Long-term orientation & 40.9 & 29.5 & 11.4\\
    Civic engagement & 40.9 & 25.0 & 15.9\\
    Mental health & 40.9 & 29.5 & 11.4\\
    True friendship & 40.9 & 34.1 & 6.8\\
    Ambitious & 40.9 & 25.0 & 15.9\\
    Courageous & 40.9 & 36.4 & 4.5\\
    Environmental sustainability & 40.9 & 20.5 & 20.5\\
    Self actualization, personal growth & 40.9 & 27.3 & 13.6\\
    Responsible & 40.9 & 29.5 & 11.4\\
    Uncertainty avoidance & 38.6 & 13.6 & 25.0\\
    Biological and Physiological needs & 38.6 & 20.5 & 18.2\\
    Salvation & 38.6 & 6.8 & 31.8\\
    Clean & 38.6 & 25.0 & 13.6\\
    Asking for clarification & 36.4 & 22.7 & 13.6\\
    Duty & 36.4 & 13.6 & 22.7\\
    Forgiving & 36.4 & 15.9 & 20.5\\
    Agreement & 36.4 & 22.7 & 13.6\\
    Restraint & 36.4 & 4.5 & 31.8\\
    Approaching others & 36.4 & 15.9 & 20.5\\
    Power distance & 36.4 & 11.4 & 25.0\\
    Safety needs & 36.4 & 25.0 & 11.4\\
    Family Security & 34.1 & 20.5 & 13.6\\
    Femininity & 34.1 & 13.6 & 20.5\\
    Labor & 34.1 & 9.1 & 25.0\\
    Committing & 31.8 & 20.5 & 11.4\\
\bottomrule
    \end{tabular}
    \caption{In the Full condition, the usage rate of values ranged from $88.6\%$ to $31.8\%$. We list all 78 values along with the percentage of users that selected the value, percentage of participants that upranked the value, and percentage of participants that downranked the value.}
    \label{tab:my_label}
\end{table*}

\begin{table*}[]
    \centering

    \footnotesize
    \begin{tabular}{lrrrr}
    \toprule
    \textbf{Value} &  \textbf{Full Weight} & \textbf{Single Weight} & \textbf{$t$} & \textbf{p}\\
    \midrule
Adding humor & $0.45\pm0.17$ & $0.72\pm0.17$ & -2.21 & 0.03\\
Quality of Content & $0.36\pm0.17$ & $0.68\pm0.21$ & -2.29 & 0.03\\
Wisdom & $0.33\pm0.23$ & $0.68\pm0.22$ & -2.04 & 0.05\\
Knowledgeable People & $0.31\pm0.22$ & $0.65\pm0.23$ & -2.03 & 0.05\\
Adding factual information & $0.28\pm0.18$ & $0.67\pm0.19$ & -2.87 & 0.01\\
Cognitive needs & $0.22\pm0.20$ & $0.58\pm0.25$ & -2.24 & 0.03\\
Self-respect & $0.17\pm0.18$ & $0.59\pm0.25$ & -2.69 & 0.01\\
Freedom & $0.13\pm0.18$ & $0.57\pm0.26$ & -2.68 & 0.01\\
Committing & $0.10\pm0.16$ & $-0.29\pm0.24$ & 2.65 & 0.01\\
Inner Harmony & $0.08\pm0.21$ & $0.56\pm0.24$ & -2.96 & 0.00\\
Equality & $0.07\pm0.24$ & $0.50\pm0.26$ & -2.31 & 0.02\\
Individualism & $0.01\pm0.19$ & $0.36\pm0.29$ & -2.03 & 0.05\\
Providing input & $-0.01\pm0.19$ & $0.40\pm0.25$ & -2.52 & 0.01\\
A world at Peace & $-0.02\pm0.26$ & $0.62\pm0.25$ & -3.40 & 0.00\\
Collectivism & $-0.10\pm0.25$ & $0.40\pm0.29$ & -2.53 & 0.01\\
Offensive, Abusive, Harassing Content or Behaviors & $-0.21\pm0.19$ & $-0.55\pm0.23$ & 2.19 & 0.03\\
Adherence to Norms & $-0.30\pm0.18$ & $0.06\pm0.29$ & -2.16 & 0.03\\
Spam, Reposts, Bots & $-0.33\pm0.19$ & $-0.70\pm0.19$ & 2.62 & 0.01\\
\bottomrule
    \end{tabular}
    \caption{We observe significant value displacement for 17 values, meaning that the absolute weight placed on them decreases in the full-library condition compared to the single-value condition. We list the 18 values for which the mean weight applied changes significantly from the single value system condition to the full library condition, of which 17 are displaced. To identify these values, we conducted two-sample independent t-tests on the value weights in the full library condition and the corresponding single value-system condition, reporting those with $p$-values $< 0.05$. The table includes the value name, mean weight in the full library condition, mean weight in the single-value system condition, the $t$-statistic, and $p$-value.}
    \label{tab:displacement}
\end{table*}

\begin{table*}[ht]
\centering
\footnotesize
\begin{tabular}{p{0.28\textwidth} c c c p{0.30\textwidth}}
\toprule
\textbf{Value System}
  & \textbf{In Source}
  & \textbf{After Filtering}
  & \textbf{After Merging}
  & \textbf{Example of Dropped Value} \\
\midrule
Building Human Values into Recommender Systems: An Interdisciplinary Synthesis
  & 31 & 24 & 10
  & Accuracy (factuality): requires external fact-checking above the post level. \\
\midrule
Hofstede’s Cultural Dimensions
  & 14 & 14 & 13
  & Indulgence: dropped after merge process. \\
\midrule
Rokeach Value Survey
  & 36 & 32 & 22
  & Happiness: strongly correlated with "Cheerful." \\
\midrule
Maslow’s Hierarchy of Needs
  & 8  & 8  & 4
  & Esteem needs: dropped after merge process. \\
\midrule
Taxonomy of Community Values on Reddit
  & 29 & 9  & 6
  & Size of Community: not applicable to Twitter or to individual posts. \\
\midrule
Taxonomy of Value Co-creation on Weibo
  & 28 & 24 & 23
  & Reinitiating humor: strongly correlated with "Adding humor." \\
\midrule
\textbf{Total}
  & \textbf{146} & \textbf{111} & \textbf{78}
  & — \\
\bottomrule
\end{tabular}
\caption{We observe changes in value systems and value counts at each stage and provide examples of dropped values. Reasons for removing values include lack of applicability to the Twitter/X platform, inability to be operationalized at the post-level, or strong correlation with other values in the library.}
\label{tab:value-systems}
\end{table*}

\begin{table*}[ht]
\footnotesize
\centering
\label{tab:methods-summary}
\begin{tabular}{@{}p{3.5cm} p{12cm}@{}}
\toprule
\textbf{Value System}
  & \textbf{Method of Construction \& Validation} \\
\midrule

\addlinespace[0.8ex]
\textit{Building Human Values into Recommender Systems: An Interdisciplinary Synthesis}  
& Gathered multi‐stakeholder input through the Partnership on AI (PAI), a global non‐profit partnership of approximately 100 academic, civil‐society, industry, and media organizations. PAI surveyed $\sim$40 recommender‐system experts—each chose a system type (e.g., social media, news aggregator) and rated each value’s importance—then reconvened them in small‐group workshops to refine the value definitions. \cite{stray2022building} \\

\midrule
\addlinespace[0.8ex]
\textit{Hofstede’s Cultural Dimensions}  
& IBM surveyed organizational units twice over four years (100{,}000+ questionnaires), analyzed country‐level mean correlations, and re‐administered key items to $\sim$400 international management trainees (30 countries), finding strong correspondence with the original IBM database. \cite{hofstede1984culture}\\

\midrule
\addlinespace[0.8ex]
\textit{Rokeach Value Survey}  
& Terminal values selected from several hundred sources (literature, personal experience, a representative city sample of $N=100$, and graduate students); instrumental values drawn from Anderson’s 555‐word trait list (itself derived from Allport \& Odbert’s 18{,}000 trait names). \cite{rokeach1973nature}\\

\midrule
\addlinespace[0.8ex]
\textit{Maslow’s Hierarchy of Needs}  
& Formulated as a positive theory of motivation grounded in clinical, observational, and experimental evidence, with its core derived directly from clinical practice. \cite{maslow1987maslow} \\

\midrule
\addlinespace[0.8ex]
\textit{Taxonomy of Community Values on Reddit}  
& Surveyed 212 members across 627 subreddits for open‐ended value descriptions (1{,}481 responses), then iteratively categorized them into 29 subcategories under nine top‐level themes. \cite{weld2022makes}\\

\midrule
\addlinespace[0.8ex]
\textit{Taxonomy of Value Co‐creation on Weibo}  
& Collected 570 DMO‐initiated Weibo posts and 3{,}137 responses to develop a taxonomy via empirical‐to‐conceptual analysis; validated it by coding 100 posts and 823 responses in a conceptual‐to‐empirical pass. \cite{ge2018taxonomy}\\

\addlinespace[0.8ex]
\bottomrule
\end{tabular}
\caption{The value systems featured in Alexandria are sourced from previous literature and have documented construction and validation methods.}
\end{table*}

\end{document}